# Atomic layer deposited second order nonlinear optical metamaterial for back-end integration with CMOS-compatible nanophotonic circuitry

Stéphane Clemmen[1,2,†], Artur Hermans[1,2,†], Eduardo Solano[3], Jolien Dendooven[3], Kalle Koskinen[4], Martti Kauranen[4], Edouard Brainis[2,5], Christophe Detavernier[3], Roel Baets[1,2]

1. Photonics Research Group, Department of Information Technology, Ghent University-imec,
Sint-Pietersnieuwstraat 41, 9000 Ghent, Belgium
2. Center for Nano- and Biophotonics (NB-Photonics), Ghent University, Sint-Pietersnieuwstraat 41, 9000 Ghent, Belgium
3. Department of Solid State Sciences, Faculty of Sciences, Ghent University, Krijgslaan 281/S1, 9000 Ghent, Belgium
4. Optics Laboratory, Tampere University of Technology, PO Box 692, FI-33101 Tampere, Finland
5. Physics and Chemistry of Nanostructures Group, Ghent University, Krijgslaan 281/S3, 9000 Ghent, Belgium
† Author e-mail address: sclemmen@intec.ugent.be; these authors contributed equally to the work

**We report the fabrication of artificial unidimensional crystals exhibiting an effective bulk second-order nonlinearity. The crystals are created by cycling atomic layer deposition of three dielectric materials such that the resulting metamaterial is non-centrosymmetric in the direction of the deposition. Characterization of the structures by second-harmonic generation Maker-fringe measurements shows that the main component of their nonlinear susceptibility tensor is about 5 pm/V which is comparable to well-established materials and more than an order of magnitude greater than reported for a similar crystal [1-Alloatti *et al*, arXiv:1504.00101[cond-mat.mtrl-sci]]. Our demonstration opens new possibilities for second-order nonlinear effects on CMOS-compatible nanophotonic platforms.**

*OCIS codes: (190.4400) Nonlinear optics, materials, (190.4720), Optical nonlinearities of condensed matter, (190.4350) Nonlinear optics at surfaces*

http://dx.doi.org

Second-order nonlinear optical response of materials gives rise to useful effects, including nonlinear wave mixing and the Pockels effect, with applications such as light generation in optical parametric oscillators and electro-optic modulation. Some of these applications have been miniaturized using various technologies. More recently, a goal has been to integrate optical functionalities on nanophotonic chips that are compatible with CMOS fabrication, which is the standard in micro/nanoelectronics. As a result, optical parametric oscillators [2] and fast modulators [3] have been reported. Since silicon and silicon nitride, which are the two main CMOS-compatible photonics platforms, lack a second-order nonlinearity, those realizations were based on the third-order nonlinearity or carrier effects. This resulted in a modest improvement in terms of energy consumption and efficiency over simpler second-order nonlinear devices widely used in free space nonlinear optics. Therefore, it would be highly desirable to be able to induce a second-order nonlinear response in a material otherwise lacking that property.

To date, common methods to artificially create a second-order nonlinearity include poling in silica glass [4] or polymers [5], strain [6], and plasmonic surface enhancement [7]. In addition, even materials expected to lack a second-order response may in some cases exhibit significant responses, but their origin remains unknown [8,9]. In any case, the inversion symmetry of the material structure must somehow be broken to induce a second-order nonlinear response.

In this letter, we utilize a new way of breaking the symmetry of otherwise centrosymmetric materials to induce a substantial second-order nonlinear response, as described by the second-order susceptibility $\chi^{(2)}$. We deposit very thin layers of three distinct transparent amorphous materials A, B, and C and repeat that structure many times to form a thick layer of a composite ABC material. In such a system, each interface between any two materials breaks the symmetry resulting in an effective bulk $\chi^{(2)}$ for the overall structure. Such an approach was introduced in 2014 [7] and implemented by atomic layer deposition (ALD) in 2015 [1]. However, these demonstrations reported a relatively low second-order response. Here, we demonstrate that such ABC approach can result in a significantly larger $\chi^{(2)}$, comparable to that of well-known second-order materials. Our characterization is based on second-harmonic generation (SHG) Maker-fringe measurements that allow the nonlinearity of the ABC layer to be separated unambiguously from that of the substrate. We verified that the SHG contributions of each of the 3 interfaces A-B, B-C, and C-A do not sum up to 0 as would be expected for an A-B system.

It is important to understand that our approach is well suited for integration with existing CMOS-compatible nanophotonics platforms. Indeed, the deposition method, ALD, is conformal, requires low temperature, and has been proven to integrate perfectly with existing nanophotonic circuitry [10]. Moreover, as the symmetry of the ABC structure is broken along its normal, the nonlinearity requires electric field components along the normal direction, which occurs for p-polarized light at non-normal incidence (see figure 1a). This implies that, for the case of widely used planar or rib waveguides, the nonlinearity would be the greatest for a TM-mode such as illustrated in figure 1b.

In the present proof of principle, the three materials were chosen to be (A) $TiO_2$, (B) $Al_2O_3$, and (C) $In_2O_3$. There are no detailed theories for predicting the second-order nonlinearity of the interface between any two materials. In choosing our particular materials, we therefore used the old guideline that the nonlinearity depends on the dielectric contrast between the materials [11] (see table 1). In consequence, the nonlinearities of the AB and BA interfaces are equal in magnitude but opposite in sign. By combining the three materials appropriately, however, we expect to obtain a structure where the nonlinearities of the AB, BC, and CA have a non-cancelling contribution. Each individual layer is 0.7 nm thick such that the ABC cycle is repeated many times to form thick ABC-composite layers on top of 500 μm-thick borosilicate glass substrates (Schott BOROFLOAT®33). The thickness of individual layers is an arbitrary compromise between the highest number of interfaces possible and the certainty to have a well-defined layer. Several ABC-composite crystals were made that differ only by their total thickness of 2.1, 25 and 50 nm which is limited solely by the duration of the deposition process. The deposition process started with the cleaning of the glass with $O_2$ plasma. The ALD was enhanced by using an oxygen plasma with an RF power of 200 W, a frequency of 13.56 MHz and a pulse duration of 10 s. The deposition was done by alternating pulses of the corresponding metalorganic precursor at a pressure of $6.0 \times 10^{-5}$ bar, followed by the $O_2$ plasma pulse at $1.2 \times 10^{-5}$ bar and a temperature of 120°C, constant throughout the full deposition process. Table 1 shows the precursors and growth-per-cycle for each of the three materials deposited [12-15].

| Material | Precursor | Growth nm/cycle | Bandgap eV | Refractive index |
|---|---|---|---|---|
| $TiO_2$ | TDMAT – Tetrakis (dimethylamido) titanium | 0.06 | 3.4 [15] | 2.1 [13] |
| $Al_2O_3$ | TMA - Trimethylaluminium | 0.1 | 8.8 [14] | 1.6 – 1.7 [12] |
| $In_2O_3$ | In(TMHD)3 – Tris(2,2,6,6-tetramethyl-3,5-heptanedionato) indium(III) | 0.01 | 3.7 [16] | 2.2 [16] |

Table 1: Summary of ALD parameters and optical properties. For the TDMAT and In(TMHD)3 precursors, Ar was used as a carrier gas.

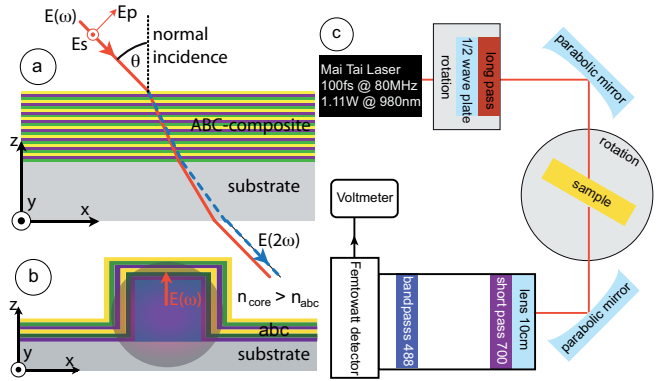

Figure 1. (a) Geometry of the ABC-composite and the incidence of either p- or s-polarized light resulting in collinear second harmonic generation. (b) Possible use of the nonlinearity via the TM-mode of a waveguiding structure (c) Maker fringes experimental setup for characterization of $\chi^{(2)}$ nonlinearity via second-harmonic generation.

| ABC-composite | | Brosilicate glass | |
|---|---|---|---|
| $n_z$ (980 nm) ordinary | 2.02 | n (980 nm) | 1.4633 |
| $n_z$ (490nm) ordinary | 2.13 | N (490 nm) | 1.4766 |
| $n_x$ (980 nm) extraordinary | 1.92 | $\chi_{xxz}^{glass}$ (m$^2$/V) | $11.6 \pm 0.8 \times 10^{-22}$ [19] |
| $n_x$ (490nm) extraordinary | 2.06 | $\chi_{zzz}^{glass}$ (m$^2$/V) | $93 \pm 14 \times 10^{-22}$ [19] |

Table 2: Parameters used for the fitting (see body of the text for more details) Left: Refractive indices of the ABC-composite measured by ellipsometry at the fundamental and second-harmonic wavelengths for ordinary and extraordinary polarization. Right: refractive index and nonlinearity of the borosilicate glass substrate.

Since the individual layers are much thinner than optical wavelengths, we can consider the multilayer as one homogeneous uniaxial material [16]. The refractive indices of the ABC-composite were measured via ellipsometry (see table 2).

The nonlinear characterization was done using the Maker-fringe technique [17] with the setup depicted in figure 1c. The source for fundamental light was a commercial Ti:sapphire laser (Mai Tai HP from Spectra-Physics) emitting 100 fs pulse at the fundamental wavelength of 980 nm and a peak power reaching 140 kW. The linear polarization of the laser beam was aligned to p-polarization with a half-wave plate and its spectrum was cleaned to suppress any spurious light at wavelengths below 800 nm. Then, the light is slightly focused on the sample using a parabolic mirror of 5 cm focal length such that the beam size (1/e$^2$) is brought from 1.2 mm to 52 ± 4 microns. The corresponding Rayleigh range is 9 mm thus leaving a relatively large tolerance for the alignment of the sample in the focal plane. After interaction with the sample, the light was collimated and the fundamental wavelength filtered out from the generated second-harmonic light. A lens was placed before the femtowatt detector (Thorlabs PDF10A) so that it can accommodate beam displacements induced by the rotation of the sample. To confirm that the detected signal is SHG and not any fluorescence, we tuned the laser wavelength so that its corresponding second-harmonic wavelength falls out of the bandpass filter and verified that the signal collected vanishes.

The results of our experiment consist of Maker fringes for our samples consisting of a blank substrate, and substrates coated on one side by 2.1, 25, and 50 nm of our ABC-composite material. The results are summarized in figures 2 and 3. The curve corresponding to the bare borosilicate glass shows the expected Maker fringes with a visibility limited in part by temporal walk-off occurring between the fundamental beam and the second-harmonic signal generated at the first interface and for the rest due to experimental imperfection leading to slightly unequal intensities at the front and back surfaces. The curve corresponding to 2.1 nm sample shows essentially the same response as the blank substrate thus providing a first indication that the air-ABC interface is not responsible for the increased SHG. The curves corresponding to thicker ABC samples show SHG increasing with thickness and fringes with reduced visibility. The reduced visibility is due to imbalance between the increased responses of the ABC layers and the unchanged response of the uncoated back surface.

To extract a value for the second-order susceptibility from these measurements, we model and fit the experimental curves assuming that SHG occurs at the ABC layer at the front interface and at the glass-air back interface and depends on the respective components of the second-order nonlinear susceptibility tensor $\chi_{xxz}^{abc}, \chi_{zxx}^{abc}, \chi_{zzz}^{abc}$ of the ABC composite and $\chi_{xxz}^{glass}, \chi_{zxx}^{glass}, \chi_{zzz}^{glass}$ of the glass substrate. These signals can be described by the following equations [18]:

$$E_{front} \propto \frac{t_{air,abc}^2 T_{abc,glass} T_{glass,air}}{N_{abc} \cos(\theta_{abc})} exp\left(i \frac{2L\omega N_{glass} \cos \theta_{glass}}{c}\right)$$
$$\times (\chi_{xxz}^{abc} \sin(2\theta_{abc}) \cos(\theta_{abc})$$
$$+ \chi_{zxx}^{abc} \sin(\theta_{abc}) \cos^2(\theta_{abc})$$
$$+ \chi_{zzz}^{abc} \sin^2(\theta_{abc}) \sin(\theta_{abc}))$$

$$E_{back} \propto \frac{t_{air,abc}^2 t_{abc,glass}^2 T_{glass,air}}{N_{glass} \cos(\Theta_{glass})} exp\left(i\frac{2L\omega n_{glass} \cos\theta_{glass}}{c}\right)$$
$$\times (\chi_{xxz}^{glass} \sin(2\theta_{glass}) \cos(\Theta_{glass})$$
$$+ \chi_{zxx}^{glass} \sin(\Theta_{glass}) \cos^2(\theta_{glass})$$
$$+ \chi_{zzz}^{glass} \sin^2(\theta_{glass}) \sin(\Theta_{glass}))$$

The generated fields depend on parameters defined at the fundamental frequency ω (lower case letters) and second-harmonic frequency 2ω (capital letters), such as the Fresnel transmission coefficients $t_{i,j}$ and $T_{ij}$ at each interface, the propagation angles θ and Θ of the beams in the ABC layer and in the glass, as well as the refractive indices of the glass substrate $n_{glass}$ and $N_{glass}$. As our model does not account for the birefringence of the material, we also neglected it in the Fresnel transmission coefficients and set the indices as the average of the ordinary and extraordinary indices (n=1.97, N=2.1). The thickness $L$ and dispersion of the glass substrate are responsible for the period of the Maker fringes as the intensity is given by I = $c_1$ | $E_{front}$ + $E_{back}$|² where $c_1$ is a proportionality constant. The fitting procedure contains three real-valued free parameters: $c_1$, $\chi_{zzz}^{abc}$ and $(2\chi_{xxz}^{abc} + \chi_{zxx}^{abc})$. Other parameters used for the fitting are summarized in table 2. Assuming that the SHG originates from an effective bulk nonlinearity of the ABC layer, the respective bulk second-order susceptibility is obtained by dividing measured surface-type signals by the thickness D of the ABC layer so that $\chi_{zzz} \equiv \frac{\chi_{zzz}^{abc}}{D}$ and $A_{zx} \equiv \frac{(2\chi_{xxz}^{abc} + \chi_{zxx}^{abc})}{D}$. The small discrepancy between experimental and fitting curve at small angles of incidence may be due to multiple reflections between the front and back surfaces, which our model does not account for. Our characterization also suffers from imperfections that manifest themselves in the non-diagonal components $(2\chi_{xxz}^{abc} + \chi_{zxx}^{abc})$, whose values for different samples vary by a factor of two. Note, however, that these components are particularly sensitive to the quality of the fit for small angles of incidence, whereas the diagonal component is relatively more important for large angles of incidence, where the fit is very good.

From this fitting procedure, we deduce the nonlinearity to be $\chi_{zzz}$ = 6.1 ± 0.4 pm/V ($\chi_{zzz}$ = 6.0 ± 0.8 pm/V) and $A_{zx}$ = 0.78 ± 0.07 pm/V ($A_{zx}$ = 1.44 ± 0.16 pm/V) for the 50 nm (25 nm) thick samples. Clearly, the main diagonal tensor component $\chi_{zzz}$ is significantly larger than the value of 0.26 pm/V reported before for a similar nanocomposite [1]. Assuming that Kleinman symmetry is satisfied, we can impose $\chi_{xxz} = \chi_{zzx}$ and then deduce also an order of magnitude value for $\chi_{zzx} \approx 0.35 \pm 0.15$ pm/V.

To gain further insight on the relative strengths between the diagonal and non-diagonal components, we also measured SHG from our 50-nm thick sample as a function of the polarization of the fundamental beam at a fixed incident angle. Figure 3 shows that the SHG signal vanishes almost perfectly for s-polarized incident light, which confirms that the non-diagonal components of the nonlinear tensor are indeed much weaker than the diagonal one. Further studies are needed to obtain more precise values for the non-diagonal tensor components of our ABC composite.

To further demonstrate that the origin of the effective bulk nonlinearity is the broken centrosymmetry resulting from the ABC structure, we have acquired additional Maker fringes (see figure 4) from a structure where two ABC samples were brought together. The reference curve (crosses) corresponds to SHG from two blank glass substrates thus showing a weak response. The level of SHG is almost identical to the reference curve presented in figure 2 but presents fringes with a shorter period and reduced visibility as a consequence of the doubled thickness of the substrate. In particular, the temporal walk-off becomes more significant because of the thicker sample. The second curve (squares) is obtained using two identical ABC samples, coated on one side of the substrate. The two samples face each other on their uncoated sides so that the interfaces producing SHG are the two ABC layers while the glass-glass interface doesn't produce SHG. This results in strong SHG with fringes exhibiting similar visibility as the blank sample. However, the visibility is still reduced for the reasons mentioned above. Finally, the last curve (circles) corresponds to those two same ABC-coated substrates but facing each other on their coated side. The nonlinear material has thus a restored symmetry as its cycling structure is now ABC…ABC-CBA…CBA and it is expected to result in negligible SHG. Indeed, while the SHG contributions from the air-glass interfaces of course remain, the curves indicate that the contributions from the ABC layers vanish to a large extent.

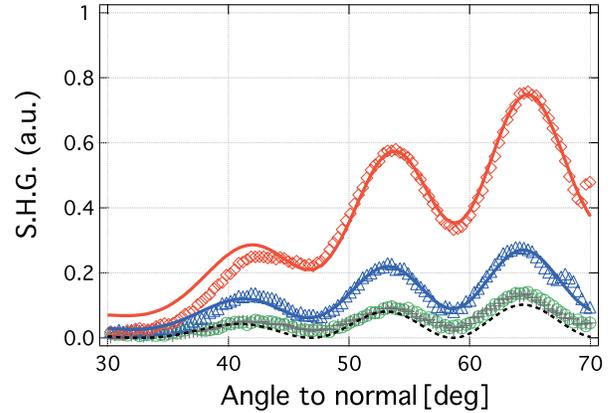

Figure 2 : Experimental data (markers) and fitting curves (bold lines) of the second harmonic generation as a function of the incidence angle for 50 nm (diamonds), 25 nm (triangles), 2.1 nm (circles) and 0 nm (crosses) thick ABC composite coating deposited on the front surface of a borofloat wafer.

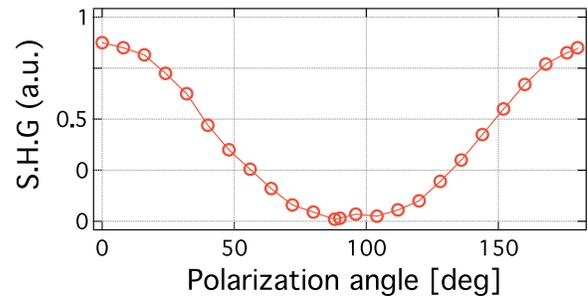

Figure 3. Variation of the S.H.G. with the incident polarization angle.

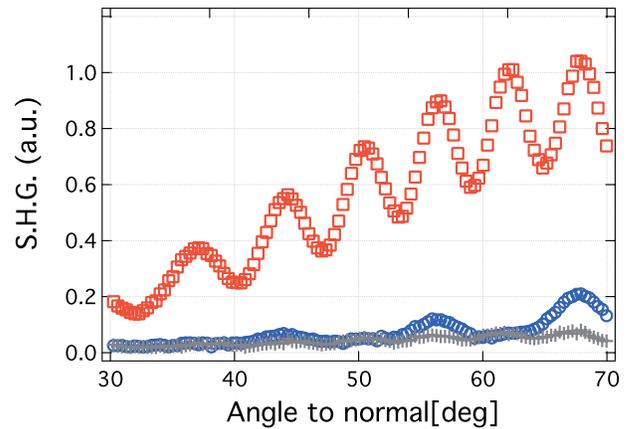

Figure 4. Comparison of SHG originating from 2 parts-samples: ABC-glass+glass-CBA (red squares), glass-CBA+ABC-glass (blue circles), glass+glass (grey crosses).

We should note that the samples of the present study were far from being optimized. In addition, our experimental setup was not yet optimized for the most precise measurements. While this affects mainly the weaker non-diagonal tensor components, both the diagonal and non-diagonal components are partly coupled through our fitting procedure. In order to take this remaining uncertainty into account, we believe that it is safe to state that the value of the dominant component is $\chi_{zzz}$ is $5 \pm 2$ pm/V.

Nevertheless this value for the main tensor component is greater by more than an order of magnitude than the 0.26 pm/V reported by Alloatti *et al.* [1] for a similar system. This might be due the fact that in that study, the interference between SHG from the front and back surfaces is not accounted for, i.e., the Maker fringes are not resolved, despite the fact that the Rayleigh range is much thicker than the sample. This is likely to result in an underestimation of the nonlinearity if the Maker-fringe maximum is missed. Moreover, one of the constituent materials of the ABC composite is different. Finally, the ALD process itself may be different, which according to Alloatti *et al.* [1] may also affect the overall quality of the composite.

In conclusion, we have demonstrated an artificial nonlinear material relying on the principle of surface induced symmetry breaking. We believe that this new class of nonlinear material is promising as our proof of principle indicates a second-order nonlinearity reaching 5±2 pm/V for its main tensor component. We believe that such a second order nonlinearity could be used in combination with nanophotonic waveguides based on CMOS-compatible material that lack significant second-order nonlinearity. Furthermore, the possibilities to increase the nonlinearity of the ABC-composite are numerous ranging from thinner individual layers to optimization of the contrast between the materials involved [11].

**Funding.** S.C., R.B., A.H., and J.D. would like to thank the ERC-InSpectra advanced grant and FWO-Vlaanderen for funding support. K.K. and M.K. acknowledge Tampere University of Technology for Optics and Photonics Strategic Funding. K.K. acknowledges the Vaisala Foundation for a Fellowship.